\documentclass[draftclsnofoot,onecolumn, 12pt]{IEEEtran}
\usepackage{graphicx}
\usepackage{graphicx}
\usepackage{amsmath}
\usepackage{amsfonts}
\usepackage{amssymb, color, bm}
\let\labelindent\relax

\usepackage{algorithm}
\usepackage{algorithmicx}
\usepackage[noend]{algpseudocode}
\usepackage{comment}

\usepackage[noadjust]{cite}

\usepackage{fancyhdr}
\usepackage{enumitem} 

\begin{document}

\title{Deep Reinforcement Learning-Aided Random Access}
\author{Ivana Nikoloska, \emph{Student Member, IEEE}, and Nikola Zlatanov, \emph{Member, IEEE} \thanks{Ivana Nikoloska and Nikola Zlatanov are with the Department of Electrical and Computer Systems Engineering, Monash University, Melbourne, Australia. Emails: ivana.nikoloska@monash.edu, nikola.zlatanov@monash.edu} }
\maketitle

\begin{abstract}
 We consider a system model comprised of an access point (AP) and $K$ Internet of Things (IoT) nodes that sporadically become active in order to send data to the AP. The AP is assumed to have $N$  time-frequency resource blocks that it can allocate to the  IoT nodes that wish to send data, where $N < K$. The main problem is how to allocate  the $N$ time-frequency resource blocks to the IoT nodes in each time slot such that the average packet rate is maximized. For this problem,  we propose a deep reinforcement learning (DRL)-aided random access (RA) scheme, where  an intelligent DRL agent at the AP learns to predict the  activity of the IoT nodes in each time slot and grants  time-frequency resource blocks to the IoT  nodes predicted as active. Next, the IoT nodes that are missclassified as non-active by the DRL agent, as well as unseen or newly arrived nodes in the cell, employ the standard RA scheme in order to obtain  time-frequency resource blocks. We leverage expert knowledge for faster training of the DRL agent. Our numerical results show significant improvements in terms of average packet rate when the proposed DRL-aided RA scheme is implemented  compared to the existing solution used in practice, the standard RA scheme. 
\end{abstract}


\section{Introduction}\label{Sec-Intro}
Given its ubiquitous coverage, the 5-th Generation of cellular networks (5G) has great potential to support diverse wireless technologies. These wireless technologies are expected to be crucial in next generation smart cities, smart homes, automated factories, automated health management systems, and many other applications, some of which can not even be foreseen today \cite{7073806}. The heterogeneous ecosystem of applications will result in different and often conflicting demands on the 5G radio and, as a result, the air-interface must be capable of supporting both high and low data rates, mobility, (ultra) low latency, as well as many different types of Quality of Service (QoS). In order for this QoS diversity to be achieved, improved medium access control (MAC) protocols have to be developed. As the state of the art MAC protocols for cellular networks have been designed and optimized to support primarily Human-to-Human (H2H) communication, these MAC protocols are not optimal for a massive number of Internet of Things (IoT) devices, due to the unique characteristics of the IoT traffic.

In IoT networks, communication devices can operate autonomously with little or no human intervention. The amount of data which is generated by the IoT nodes is usually small and the communication activity of these devices is heterogeneous \cite{shober}. The heterogeneous communication activity of the IoT nodes makes any attempt to pre-allocate network resources to each  IoT node spectrally inefficient \cite{laya2014random}. Currently,  IoT devices  gain access to the channel, and thereby transmit information, by performing  grant-based random access (RA) or grant-free RA \cite{3GPP_MAC}. In grant-based RA, the nodes attempting to access the channel have to first obtain an access grant from an Access Point (AP) through a four-way handshake procedure \cite{3GPP_MAC}. This ensures that the user has exclusive rights to the channel if granted access, thus avoiding any potential collisions, at the expense of large latency and signalling overhead. In grant-free RA, the data is piggy-backed on the first transmission itself along with the required control information, in order to reduce the access latency. 
 However, both schemes   suffer from massive packet collisions as the number of IoT nodes  requiring access increases. Packet collisions increase latency and energy inefficiency, since collided packets need to be retransmitted,  and require heavy exchange of signalling messages. As a result, packet collisions in massive IoT networks can  easily
become a bottleneck.

One promising research direction for MAC related problems in wireless communication is machine learning. Reinforcement Learning (RL) is one of many machine learning paradigms, where agents mimic the human learning process and learn optimal strategies by trial-and-error interactions with the environment \cite{sutton1998reinforcement}. RL has been implemented in the development of MAC schemes for cognitive radios \cite{bkassiny2011distributed}, where the authors have developed a RL based MAC scheme which allows each autonomous cognitive radio to distributively
learn its own spectrum sensing policy. In \cite{chu2012aloha}, the authors add intelligence to sensor nodes to improve the performance of Slotted ALOHA. In addition, solving MAC problems with multi agent Deep Reinforcement Learning (DRL) has been proposed in \cite{naparstek2017deep},\cite{yu2017deep},\cite{wang2017deep}. Specifically, in \cite{naparstek2017deep}, the authors propose a DRL MAC protocol for wireless networks in which multiple agents learn when to access the channel. A DRL MAC protocol for wireless networks in which several different MAC protocols co-exist has been studied in \cite{yu2017deep}. The authors of \cite{wang2017deep} have proposed a multi-agent DRL-based MAC scheme for wireless sensor networks with multiple frequency channels. Another distributed MAC scheme has been investigated in \cite{bello2014application}, where authors embed learning mechanisms to the IoT nodes in order to control IoT traffic and consequently reduce its impact on any cellular network. In \cite{moon2017access}, \cite{moon2017reinforcement}, \cite{tello2018reinforcement}, the authors reduce the access congestion by adapting the parameters of the access class baring mechanism to different IoT traffic conditions, via DRL. A different approach is proposed in \cite{liu2017enb}, \cite{mohammed2015base}, where the authors investigate learning mechanisms to aid the MAC in IoT networks, via dynamic AP selection schemes, in order to avoid overloading a single AP. 

In spite of being highly promising, the schemes proposed in \cite{chu2012aloha}-\cite{bello2014application} do not necessarily account for the severe device constraints in terms of energy availability and computing power for running on-device optimization and inferences \cite{park2016learning}. On the other hand, the schemes in \cite{moon2017access}-\cite{tello2018reinforcement} still rely on RA as a primary access mechanism. In this paper, we propose a DRL-aided RA scheme which does not require on-device inferences at the IoT nodes and therefore is applicable to devices with computational and energy constraints. In particular, we consider an IoT network comprised of an AP and $K$ IoT nodes that sporadically become active and transmit information towards the AP. \textcolor{black}{Practical applications that subscribe to these assumptions include smart metering, temperature monitoring, air-quality monitoring, emergency reporting etc.} 
In the proposed scheme, the AP is assumed to have $N$  time-frequency resource blocks that it can allocate to the  IoT nodes that wish to send data, where $N < K$. The main problem is how to allocate  the $N$ time-frequency resource blocks to the IoT nodes in each time slot such that the average packet rate received at the AP is maximized. For this problem,  we propose a DRL-aided RA scheme, where  an intelligent DRL agent at the AP learns to predict the  activity of the IoT nodes in each time slot and grants  time-frequency resource blocks to the IoT  nodes predicted as active. Next, the IoT nodes that are missclassified as non-active by the DRL agent, as well as unseen or newly arrived nodes in the cell, employ the standard RA scheme in order to obtain  time-frequency resource blocks. In this paper, we rely on grand-based RA, however, the proposed hybrid scheme is also compatible with grant-free RA. To reduce the amount of live data which needs to be acquired from the IoT network for training the DRL agent, we propose to leverage expert knowledge from the available theoretical models in the literature. Our numerical results show that the proposed algorithm significantly increases the packet rate, and implicitly decreases the energy consumption of the IoT nodes. In addition, as the intelligence is concentrated at the AP, the IoT nodes do not need significant computational power, or energy, for the on-device inferences, and thereby the proposed scheme can be deployed in cells with generic IoT nodes that have limited computational capabilities. 

The promising results of the proposed scheme are due to the fact that the conventional RA scheme can not utilize the determinism in the nodes' activity patterns, which usually exists in practice \cite{ferdouse2017congestion}. Our proposed DRL-aided RA scheme fills in this gap. Specifically, the proposed DRL-aided RA scheme uses the DRL algorithm to learn the deterministic components of the nodes' activity patterns in order to allocate time-frequency resources. Moreover, the proposed DRL-aided RA scheme uses the conventional RA scheme to cope with the random components of the nodes' activity patterns and allocate time-frequency resources in the presence of random components.  In that sense, the proposed DRL-aided RA scheme operates in the range between the two limiting type of activity patterns. At one end of the range is the absolutely independent and identically distributed (i.i.d.) random activity pattern and at the other end of the range is the absolutely deterministic activity pattern.

The rest of the paper is organized as follows. Section~II provides the network model. Section~III presents the proposed DRL-aided RA algorithm. In Section~IV, we provide numerical evaluation, and a short conclusion concludes the paper in Section~V.

\begin{figure} 
\centering
\includegraphics[width=5.8in]{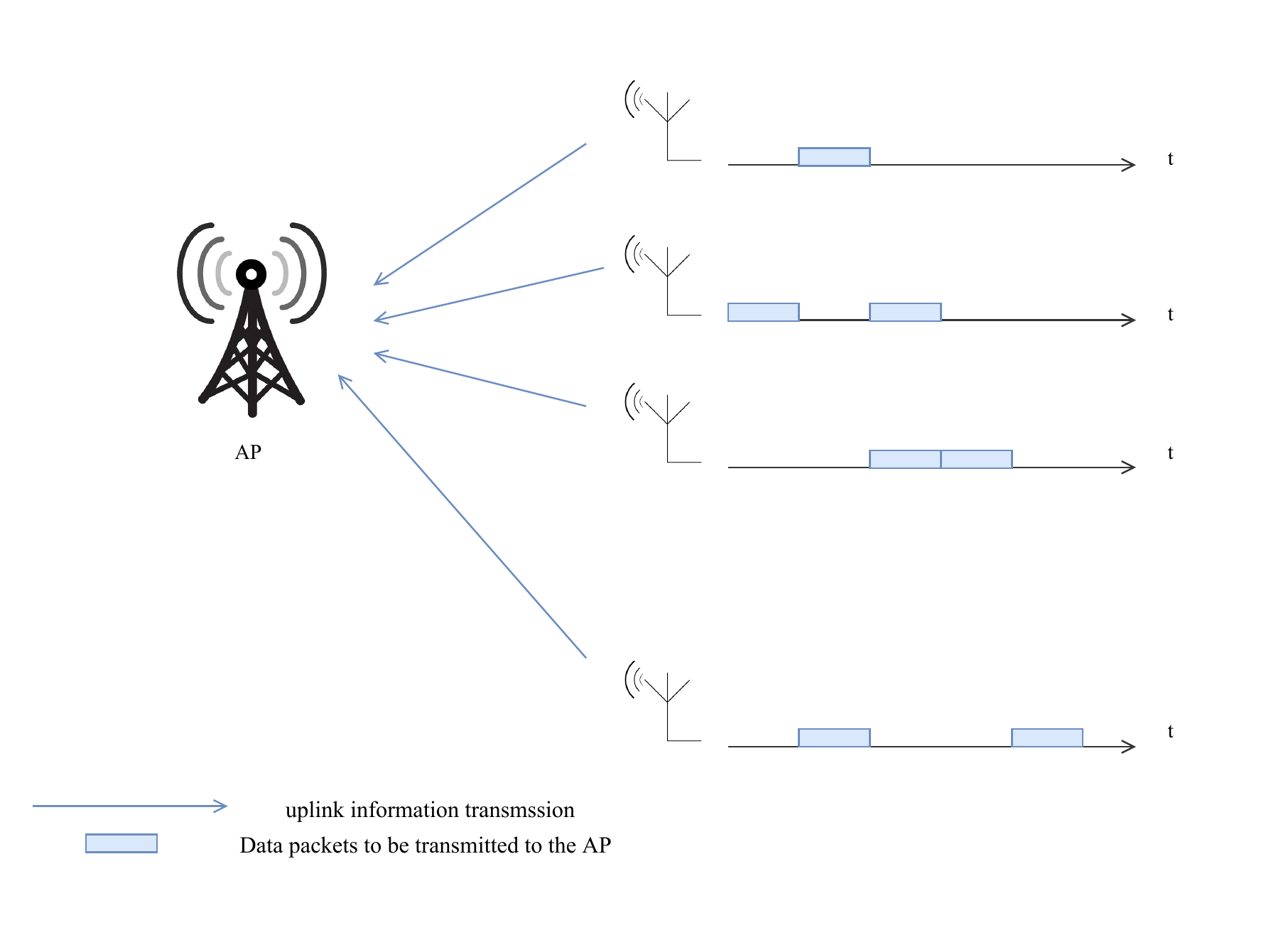}
\caption{Network model.}
\label{sys1}
\end{figure}

\section{System Model And Problem Formulation}\label{Sec-Sys}

In the following, we provide the system model and formulate  the underlying problem.

\subsection{System Model}

We consider a network comprised of $K$ IoT nodes and an AP, as illustrated in Fig.~\ref{sys1}. The locations of the IoT nodes are assumed to be fixed and not to change with time.  
The transmission time is   divided into $T$ time slots of equal duration.
At the beginning of each time slot, each IoT node sporadically becomes active  in order to sense its environment, generates a   data packet from the sensed data, and tries to transmit this data packet to the AP in the same time slot. 
In order for an  IoT node to transmit a data packet to the AP  in time slot $t$,   a dedicated time-frequency block, refereed to as resource block (RB), needs to be  allocated to the   IoT node. Without loss of generality, we assume that all nodes transmit their packets with identical data rate, which is set to one. We assume that    the AP has  $N$ RBs available in total, where $N < K$ holds.  As a result, in each time slot,  the AP needs to perform intelligent  resource allocation by allocating the available $N$ RBs to the active IoT nodes only. Otherwise, if the AP allocates a  RB to a non-active node, that RB would be wasted and the AP will not receive a packet on the corresponding RB.

\subsection{Problem Formulation}

 The AP    receives a packet from the $k$-th IoT node in time slot $t$ if the following two events occur: 
\begin{itemize}
\item the  $k$-th IoT node is active in time slot $t$,
\item the  $k$-th IoT node has been allocated a RB in time slot $t$.
\end{itemize}
Otherwise, the AP will not  receive a packet from the $k$-th IoT node in time slot $t$. 
 To model this behaviour, 
let $A_k(t)$ and $I_k(t)$
 be  binary indicators defined as
\begin{align}
A_k(t)&=\left\{
\begin{array}{ll}
1 &\textrm{if  node } k \textrm{ is active in time slot } t \\
0 &\textrm{otherwise},
\end{array}
\right.\\
I_k(t)&=\left\{
\begin{array}{ll}
1 &\textrm{if a RB has been allocated to node } k \textrm{ in time slot } t \\
0 &\textrm{otherwise}.
\end{array}
\right.
\end{align}
Using these binary indicators, we can obtain   the  average packet rate, denoted by $R$, as
\begin{align}\label{eq-rate}
R= \frac{1}{T} \frac{1}{K} \sum_{t=1}^T \sum_{k=1}^K  A_k(t)I_k(t) .
\end{align}
Our aim is to maximize the average packet rate by solving the following    optimization problem 
\begin{align}\label{eq-max-prob}
&\max_{I_k(t)} \frac{1}{T} \frac{1}{K} \sum_{t=1}^T \sum_{k=1}^K  A_k(t)I_k(t) \nonumber\\
&C1: I_k(t)\in\{0,1\}\nonumber\\
&C2: \sum_{k=1}^K I_k(t)\leq N ,
\end{align}
where the last constraint follows since the AP has $N$ RBs in total in each time slot. If $A_k(t), \forall k,t$ are known, then the optimal solution of \eqref{eq-max-prob} is known  and is  $I_k(t)=1$ if  $A_k(t)=1$  until all $N$ RBs are used up. However, the main problem is that   $A_k(t)$   is unknown  in practice and needs to be estimated. 
 As a result,   a practical algorithm that provides the optimal solution to  the  maximization problem in \eqref{eq-max-prob} is difficult in general. Hence, our aim in this paper is to propose a suboptimal but a practical solution to the  recourse allocation problem in \eqref{eq-max-prob}, which provides good performance. 

\section{Proposed Solution}
In the following, we discuss the existing solution used in practice and propose our solution.

\subsection{Existing Solution: The  RA Scheme}\label{Sec-Algo-RA}

The existing practical suboptimal solution to the resource allocation  problem in \eqref{eq-max-prob} is the conventional RA scheme \cite{3GPP_MAC}. In the RA scheme, each of the $K$ IoT nodes has an identical set of $M$ orthonormal sequences. At the start of each time slot, each active node selects uniformly at random a single orthonormal sequences from its set, and uses that sequences to transmit information to the AP via  a dedicated control channel. This transmission, if successful,  informs the AP that the considered node is active in the current time slot and thereby needs to be allocated a RB.  The AP is able to receive this information from a given active node correctly if no other active node  has selected the same orthonormal sequence as the considered node. Otherwise, if two or more active nodes have selected the same orthonormal sequence,  collisions occur  and the AP is not able to receive the information from these nodes correctly. As a result, the AP will not know that these nodes will be active in time slot $t$, and consequently the AP will not grant  RBs to these nodes. In addition, the RA scheme also fails if the AP does not have enough RBs to grant to all active nodes which have successfully completed the RA procedure and informed the AP that they are active. 
 
The average packet rate achieved by the RA scheme is   given by
\begin{align}\label{eq-rate-RA}
R_{\rm RA} =  \frac{1}{T} \frac{1}{K} \sum_{t=1}^T \text{min} \left\{ \left(1- \frac{1}{M}  \right)^{K^a(t)- 1}, \frac{N}{K^a(t) \left(1- \frac{1}{M}  \right)^{K^a(t)- 1}} \right\},
\end{align}
where $(1-1/M)^n$   is the probability that collisions will not occur if $n$ nodes are active, and   $K^a(t)$ denotes the number of active nodes at time slot $t$ found as
\begin{align}
    K^a(t) = \sum_{k=1}^K A_k (t).
\end{align}

\subsection{Proposed Solution: The DRL-Aided RA Scheme}\label{Sec-Algo}

In the proposed DRL-aided RA scheme, the allocation of the $N$ RBs is conducted in two consecutive phases. In the first phase, $N_1$ RBs are allocated by the AP using the DRL algorithm presented below, where $N_1\leq N$. Next, in the following   phase, $N_2=N-N_1$ RBs are allocated by the AP using the conventional RA scheme, described in Sec.~\ref{Sec-Algo-RA}. To this end, the AP is assumed to host a DRL agent that learns to predict which nodes will be   active   in a given time slot $t$.  
 The learning and resource allocation process of the DRL agent, which is repeated in each time slot, is as follows:


\begin{itemize}
\item We define a state in time slot $t$, denoted by $\mathcal{S}_t$. The state $\mathcal{S}_t$ represents a set comprised of the nodes which have been active during the previous $t_h$ time slots, i.e., $t-t_h,...,t-1$, where $t_h$ denotes the history that the agent "remembers". 


\item Based on the state in time slot $t$, the DRL agent produces an output set,  denoted by $\mathcal{A}_t$ and referred to as the action,  comprised of the nodes that have been predicted to be active in time slot $t$  by the DRL agent. 

\item Based on the set $\mathcal{A}_t$, the following RBs allocations occur:

    \begin{itemize}\renewcommand{\labelitemiv}{$\circ$}
    \item If $|\mathcal{A}_t|\leq N_1$, each  node in the set $\mathcal{A}_t$ is allocated a  RB.
    \item If $|\mathcal{A}_t|> N_1$, then $N_1$ nodes from the set $\mathcal{A}_t$ are selected uniformly at random and each of these nodes is allocated a RB. 
    \end{itemize}

\item Based on the allocations of RBs to the nodes predicted as active by the DRL agent, the following occurs in time slot $t$

    \begin{itemize}\renewcommand{\labelitemiv}{$\circ$}
    \item If a node has been correctly predicted as active,  and thereby granted a  RB, the node  transmits its data packet  to the AP on the corresponding RB, and the AP receives this data packet correctly. Consequently, the DRL agent will classify this node as correctly predicted.
    \item If a node has been miss-predicted as active by the DRL agent and thereby has been granted a  RB, this node stays silent since it is inactive. As a result, the AP will not receive a packet on the corresponding RB. Consequently, the DRL agent will classify this node as erroneously predicted.
    \item If a node has been correctly predicted as inactive  and thereby has not been granted a  RB,    the  node stays silent. 
    \item If a node has been miss-predicted as inactive and thereby has not been granted a  RB, the miss-predicted active node attempts to obtain a  RB using the conventional RA scheme, as described in Sec.~\ref{Sec-Algo-RA}. Thereby, the node  selects uniformly at random a single orthonormal sequence  from its set, and uses that sequence  to inform the AP, via the control channel,   that the considered node is active and has been miss-predicted in the current time slot. The AP listens to the control channel and detects the nodes which have been miss-predicted as inactive. The AP can detect only those nodes which have selected  a unique orthonormal sequence. The other nodes cannot be detected due to   collisions of their packets, as explained in Sec.~\ref{Sec-Algo-RA}.
    \end{itemize}

\item Next, based on the observations, the AP constructs the set $\mathcal{S}_t$ by including the following nodes
    \begin{itemize}\renewcommand{\labelitemiv}{$\circ$}
    \item the nodes which the DRL predicted as active and from which the AP received a packet on the corresponding allocated RB.
    \item the nodes which the AP detected as active on the control channel via the RA scheme.
    \end{itemize}

\item Based on the  observation, the AP computes a reward in time slot $t$, denoted by  $r_{t}$, which is equal to $K$, if all nodes are correctly predicted in time slot $t$ by the DRl agent, or $0$ otherwise.

\item The system transitions to the next time slot and the whole process described above is repeated.


\end{itemize}

\subsubsection{Implementation of the DRL Agent}

The DRL agent is implemented as a deep neural network located at the AP. This deep neural network,  in time slot $t$, has the set $\mathcal{S}_t$ as input and produces as   outputs $L=K!$ values, denoted  by $Q_e(  \mathcal{S}_{t}|\mathcal{A}_1;\bm{\theta})$,  $Q_e( \mathcal{S}_{t}|\mathcal{A}_2;\bm{\theta})$, ..., $Q_e( \mathcal{S}_{t}|\mathcal{A}_L;\bm{\theta})$, where $Q_e(  \mathcal{S}_{t}|\mathcal{A}_l;\bm{\theta})$ is an estimated  average reward the agent will receive in the future if the agent predicts that the set of active nodes in time slot $t$ is $\mathcal{A}_l$, for $l=1,2...,L$, and $\bm{\theta}$ is a vector comprised  of the weights of the neural network.
  Next, the agent chooses that set $\mathcal{A}$ which corresponds to the largest output value of the neural network, i.e.,    
\begin{align}
\mathcal{A}=\underset{ \mathcal{A}}{\textrm{argmax }}  Q_e( \mathcal{S}_{t}|\mathcal{A};\bm{\theta}) .
\end{align}
The function  $Q_e(\mathcal{S}_{t}|\mathcal{A};\bm{\theta})$ obtained at the output of the neural network is an estimate of the function $Q(\mathcal{S}_{t}|\mathcal{A})$,  which is known as the discounted average award. The discounted average award function is defined as
\begin{align}
Q(\mathcal{S}_{t}|\mathcal{A}) & =E\left\{\sum_{k=t}^{\infty}  \gamma^{k-t} r_{t}\Big|\mathcal{S}_{t},\mathcal{A}\right\} 
& \xrightarrow{} Q(\mathcal{S}_{t}|\mathcal{A}) + \alpha \left[ r + \gamma \,  \underset{\mathcal{A}}{\text{argmax}} \, Q(\mathcal{S}_{t+1}|\mathcal{A}) - Q(\mathcal{S}_{t}|\mathcal{A}) \right],
\end{align}
where $0\leq \gamma\leq 1$ is referred to as the discount factor. In order for the neural network to produce output functions $Q_e(\mathcal{S}_{t}|\mathcal{A};\bm{\theta})$ that are estimates of $Q(\mathcal{S}_{t}|\mathcal{A})$, the Bellman equation is used and thereby the weights in the  neural network $\bm{\theta}$ are optimized such that the following mean squared error is minimized
\begin{align}\label{loss_Q}
\mathcal{L}=E\left\{\left(r + \gamma \,  \underset{\mathcal{A}}{\text{argmax}} \, Q_e(\mathcal{S}_{t+1}|\mathcal{A};\bm{\theta})- Q_e(\mathcal{S}_{t}|\mathcal{A};\bm{\theta})  \right)^2\right\}.
\end{align}
The above minimization of the mean square error can be implemented iteratively via a stochastic gradient descent (or a variant), where in each iteration  the weights in the  neural network $\bm{\theta}$ are updated according to
\begin{align} \label{update_rule_weights}
    \bm{\theta} \leftarrow \bm{\theta} - \eta \nabla_{\bm{\theta}} \mathcal{L}.
\end{align}

 The reader is kindly referred to \cite{sutton1998reinforcement} and \cite{mnih2015human}, the references therein and the references in this paper for further details.



\subsubsection{Training the DRL Agent Using Expert Knowledge}\label{Improve-Algo}
In order for the training process of the agent to be succesful, the agent needs to obtain a sufficient number of "live" training data samples from the interaction between the AP and  the IoT nodes. Let the live data sample at time slot $t$ be  obtained  as per Subsection~\ref{Sec-Algo}. In practice, the acquisition of sufficient number of live samples can be impractical and ultimately prohibitive, due to the excessively long amount of time required to acquire the data. In these cases, transfer learning can be leveraged in order to accelerate the training process. Transfer learning is a recent trend in the ML community  where available prior knowledge about the considered problem stemming from theoretical models is embedded in the neural networks \cite{8451064}-\cite{trans}.   Transfer learning   dramatically reduces the number of live data samples that are needed for the training process to be successful. 

The IoT networks research community has provided many theoretical models for the activity of the IoT nodes, such as those in \cite{shober}, \cite{madueno2016reliable}. We choose to leverage the model in \cite{madueno2016reliable}, where the authors use a Coupled Markov Modulated Poisson Process (CMMPP) to model the activity of the nodes in a IoT cell theoretically. The CMMPP model captures both regular and alarm reporting, as well as the correlated activity behavior among the nodes.   Hence, we use the CMMPP model in \cite{madueno2016reliable} to train the DRL agent at the AP. To this end, we first synthesize artificial activity patterns of the IoT nodes, according to the CMMPP model, with which we train the DRL agent.  Thereby, during the training process, in each time slot, the DRL agent  learns to predict the active nodes  from artificial activity patterns as per Subsection~\ref{Sec-Algo}.  Once the prior knowledge has been transferred, i.e., the DRL agent has been trained using the artificial activity patterns, the  DRL-RA scheme starts using the    actual live samples from the IoT nodes. 
%

\subsubsection{Average Packet Rate}
In the following, we derive the average packet rate of the proposed DRL-aided RA  scheme.
To this end, let $\epsilon^{a} (t)$ and $\epsilon^{p} (t)$ denote the probability that a node is predicted as active and inactive at time slot $t$, respectively. Let $\epsilon^{a}_{cor.} (t)$ and $\epsilon^{p}_{mis.} (t)$ denote the probability that a node is correctly predicted as active and mispredicted as inactive at time slot $t$, respectively. Finally, let $\epsilon^a_{cor., noRBs}(t)$ denote the probability that a node is correctly predicted as active at time slot $t$, but the AP does not have any RBs left to allocate to the node. For the DRL-aided RA scheme, in each time slot, $K \epsilon^{p}_{mis.} (t)$ nodes that are misclassified as inactive, and $K \epsilon^a_{cor., noRBs}(t)$ nodes that are correctly classified as active but the DRL agent does not have enough any RBs left to allocate to them, will attempt the RA procedure in order to obtain one of $N_2$ RBs. 
The rate of the proposed algorithm is thus given by 
\begin{align}\label{prob_RA}
R_{\rm DRL+RA} = \frac{1}{T} \frac{1}{K} \sum_{t=1}^{T} \text{min} \left\{ \left(1- \frac{1}{M}  \right)^{V^a(t) - 1},  \frac{N_2}{V^a(t) \left(1- \frac{1}{M}  \right)^{V^a(t) - 1}} \right\},
\end{align}
where $V^a(t) = K \epsilon^{p}_{mis.} (t) + K \epsilon^a_{cor., noRBs}(t)$. To find $V^a(t)$, we need to calculate the number of nodes which have not been granted an RB, in spite of being correctly classified as active $K \epsilon^a_{cor., noRBs}(t)$. To do so, let $^{x} \mathbf{C}_{y}$ denote the binomial coefficient, defined as
\begin{align}\label{choose}
^{x} \mathbf{C}_{y} = \frac{x!}{y! (x-y)!}.
\end{align}
We can distinguish the following cases: $K \epsilon^{a}_{mis.}(t) >  N_1$ and $K \epsilon^{a}_{cor.}(t) >  N_1$; $K \epsilon^{a}_{mis.}(t) >  N_1$ and $K \epsilon^{a}_{cor.}(t) <  N_1$; $K \epsilon^{a}_{mis.}(t) <  N_1$ and $K \epsilon^{a}_{cor.}(t) >  N_1$; and $K \epsilon^{a}_{mis.}(t) <  N_1$ and $K \epsilon^{a}_{cor.}(t) <  N_1$. In the first case, when $K \epsilon^{a}_{mis.}(t) >  N_1$ and $K \epsilon^{a}_{cor.}(t) >  N_1$, the following can occur:
\begin{itemize}
    \item {$N_1$ of the $K \epsilon^{a}_{mis.}(t)$ misclassified nodes are granted all $N_1$ RBs and none of the $K \epsilon^{a}_{cor.}(t)$ correctly classified nodes get an RB, which occurs with probability 
\begin{align}
    \frac{^{K \epsilon^{a}_{cor.}(t)} \mathbf{C}_{0}  \times ^{K \epsilon^{a}_{mis.}(t)} \mathbf{C}_{N_1} }{^{K \epsilon^{a}(t)} \mathbf{C}_{N_1}},
\end{align}    
    }
    \item {$N_1-1$ of the $K \epsilon^{a}_{mis.}(t)$ misclassified nodes are granted $N_1-1$ RBs and one of the $K \epsilon^{a}_{cor.}(t)$ correctly classified nodes gets an RB, which occurs with probability 
\begin{align}
    \frac{^{K \epsilon^{a}_{cor.}(t)} \mathbf{C}_{1}  \times ^{K \epsilon^{a}_{mis.}(t)} \mathbf{C}_{N_1 - 1} }{^{K \epsilon^{a}(t)} \mathbf{C}_{N_1}},
\end{align}    
    }
    \item {$N_1-2$ of the $K \epsilon^{a}_{mis.}(t)$ misclassified nodes are granted $N_1-2$ RBs and two of the $K \epsilon^{a}_{cor.}(t)$ correctly classified nodes get an RB, which occurs with probability 
\begin{align}
    \frac{^{K \epsilon^{a}_{cor.}(t)} \mathbf{C}_{2}  \times ^{K \epsilon^{a}_{mis.}(t)} \mathbf{C}_{N_1 - 2} }{^{K \epsilon^{a}(t)} \mathbf{C}_{N_1}},
\end{align}    
    }
    \vdots\\
    \item {None of the $K \epsilon^{a}_{mis.}(t)$ misclassified nodes are granted RBs and $N_1$ of the $K \epsilon^{a}_{cor.}(t)$ correctly classified nodes get an RB, which occurs with probability 
\begin{align}
    \frac{^{K \epsilon^{a}_{cor.}(t)} \mathbf{C}_{N_1}  \times ^{K \epsilon^{a}_{mis.}(t)} \mathbf{C}_{0} }{^{K \epsilon^{a}(t)} \mathbf{C}_{N_1}}.
\end{align}    
    }
\end{itemize}
Thereby, when $K \epsilon^{a}_{mis.}(t) >  N_1$ and $K \epsilon^{a}_{cor.}(t) >  N_1$, $K \epsilon^a_{cor., noRBs} (t)$ is given by
\begin{align}\label{noRBs1}
K \epsilon^a_{cor., noRBs} (t) &= K \epsilon^{a}_{cor.}(t) \frac{^{K \epsilon^{a}_{cor.}(t)} \mathbf{C}_{0}  \times ^{K \epsilon^{a}_{mis.}(t)} \mathbf{C}_{N_1} }{^{K \epsilon^{a}(t)} \mathbf{C}_{N_1}} \nonumber \\
&+ (K \epsilon^{a}_{cor.}(t)- 1) \frac{^{K \epsilon^{a}_{cor.}(t)} \mathbf{C}_{1} \times ^{K \epsilon^{a}_{mis.}(t)} \mathbf{C}_{N_1 - 1} }{^{K \epsilon^{a}(t)} \mathbf{C}_{N_1}} \nonumber \\
&+ (K \epsilon^{a}_{cor.}(t) - 2) \frac{^{K \epsilon^{a}_{cor.}(t)} \mathbf{C}_{2} \times ^{K \epsilon^{a}_{mis.}(t)} \mathbf{C}_{N_1 - 2} }{^{K \epsilon^{a}(t)} \mathbf{C}_{N_1}} \nonumber \\
&\vdots \nonumber \\
&+ (K \epsilon^{a}_{cor.}(t) - N_1) \frac{^{K \epsilon^{a}_{cor.}(t)} \mathbf{C}_{N_1} \times ^{K \epsilon^{a}_{mis.}(t)} \mathbf{C}_{0} }{^{K \epsilon^{a}(t)} \mathbf{C}_{N_1}}.
\end{align}
By extending this analysis to the other three cases, we obtain $K \epsilon^a_{cor., noRBs}(t)$ as 
\begin{align}\label{noRBs}
K \epsilon^a_{cor., noRBs} (t) &= n_1 \frac{^{K \epsilon^{a}_{cor.}(t)} \mathbf{C}_{n_2}  \times ^{K \epsilon^{a}_{mis.}(t)} \mathbf{C}_{N_1 - n_2} }{^{K \epsilon^{a}(t)} \mathbf{C}_{N_1}} \nonumber \\
&+ (n_1- 1) \frac{^{K \epsilon^{a}_{cor.}(t)} \mathbf{C}_{n_2 + 1} \times ^{K \epsilon^{a}_{mis.}(t)} \mathbf{C}_{N_1 - n_2 - 1} }{^{K \epsilon^{a}(t)} \mathbf{C}_{N_1}} \nonumber \\
&+ (n_1 - 2) \frac{^{K \epsilon^{a}_{cor.}(t)} \mathbf{C}_{n_2 + 2} \times ^{K \epsilon^{a}_{mis.}(t)} \mathbf{C}_{N_1 - n_2 - 2} }{^{K \epsilon^{a}(t)} \mathbf{C}_{N_1}} \nonumber \\
&\vdots \nonumber \\
&+ m_1 \frac{^{K \epsilon^{a}_{cor.}(t)} \mathbf{C}_{m_2} \times ^{K \epsilon^{a}_{mis.}(t)} \mathbf{C}_{N_1 - m_2} }{^{K \epsilon^{a}(t)} \mathbf{C}_{N_1}},
\end{align}
where $^{x} \mathbf{C}_{y}$ is given by \eqref{choose}.
In (\ref{noRBs}), the constants $n_1$, $n_2$, $m_1$, and $m_2$ can be found as
\begin{align}\label{n1}
n_1=
\begin{cases}
K \epsilon^{a}_{cor.}(t) - (K \epsilon^{a}(t) - N_1), \,\, & \text{if} \,\,  K \epsilon^{a}_{mis.}(t) <  N_1,\\
K \epsilon^{a}_{cor.}(t), & \text{if} \,\, K \epsilon^{a}_{mis.}(t) \geq  N_1,
\end{cases}.
\end{align}
\begin{align}\label{n2}
n_2=
\begin{cases}
K \epsilon^{a}(t)- N_1, \,\, & \text{if} \,\,  K \epsilon^{a}_{mis.}(t) <  N_1,\\
0, & \text{if} \,\, K \epsilon^{a}_{mis.}(t) \geq  N_1,
\end{cases}.
\end{align}
\begin{align}\label{m1}
m_1=
\begin{cases}
0, \,\, & \text{if} \,\,  K \epsilon^{a}_{cor.}(t) \leq  N_1,\\
K \epsilon^{a}_{cor.}(t) - N_1, & \text{if} \,\, K \epsilon^{a}_{cor.}(t) >  N_1,
\end{cases}.
\end{align}
and
\begin{align}\label{m2}
m_2=
\begin{cases}
K \epsilon^{a}_{cor.}(t) , \,\, & \text{if} \,\,  K \epsilon^{a}_{cor.}(t) \leq  N_1,\\
N_1, & \text{if} \,\, K \epsilon^{a}_{cor.}(t) >  N_1.
\end{cases}.
\end{align}

\section{Numerical Results}\label{Sec-Num}
In this section, we compare the performance of the proposed DRL-aided RA scheme with the conventional RA scheme in \cite{3GPP_MAC}. To this end, in Section~IV-A, we first present the data sets that have been used in the simulations and the hyper parameters of the proposed algorithm are given in Section~IV-B. The numerical results are finally given in Section~IV-C.


\subsection{Data Sets}
\subsubsection{Synthetic Activity Patterns}
To demonstrate the effectiveness of the proposed scheme on different traffic types, we first generate synthetic data sets. Specifically, node $k$ is assumed to be active in time slot $t$ with probability $p_k(t)$, where
\begin{align}\label{syn_act}
p_k(t)=
\begin{cases}
1-\frac{\delta}{2}, \,\, & \text{if $t$ is even},\\
\frac{\delta}{2},  & \text{if $t$ is odd},
\end{cases}
\end{align}
where $\delta$ is a constant which controls the determinism of the activity patterns. Thereby, lower values of $\delta$ will result in a more periodic activity pattern, and as $\delta$ increases, the activity pattern becomes more random.

\subsubsection{Real-World Activity Pattern}
To demonstrate the effectiveness of the proposed scheme even further, real-world activity patterns are drawn from the publicly available data sets in \cite{dataREFIT}, \cite{dataAIR}, \cite{dataCOMBED}, and \cite{dataPOWER}. We assume that all nodes operate during the same time period and in the same IoT cell. The data sets in \cite{dataREFIT}-\cite{dataPOWER} are comprised of nodes
which have different reporting intervals \cite{shober}. In particular, the time elapsed between two consecutive data arrivals ranges from one second for some nodes up to 1 hour for others.

\subsection{Neural Network Hyper-Parameters}
To speed-up the training of the DRL agent, we split the neural network into an ensemble of neural networks, such that only subsets of the nodes are included in each network in the ensemble. All networks in the ensemble are identically trained, as described previously. The architecture of each neural network in the ensemble consists of a three-layer, fully connected neural network. The activation functions for the neurons are ReLU functions \cite{goodfellow2016deep}, given by
\begin{align}\label{eqn_num_2}
f(x)=
\begin{cases}
0, \,\, \text{for} \,\, x < 0,\\
x, \,\, \text{for} \,\, x \geq 0.
\end{cases}
\end{align}
The discount factor in (\ref{loss_Q}) is set to $\gamma=0.05$. The exploration-exploitation trade-off \cite{goodfellow2016deep} is controlled via the $\epsilon$-greedy algorithm, where $\epsilon_t$ is decreasing from $\epsilon_t=1$ to $\epsilon_t=0.01$ as
\begin{align}\label{eps}
\epsilon_{t+1} \leftarrow \epsilon_t * \epsilon_{dec.},
\end{align}
where $\epsilon_{dec} = 0.995$. Thereby, the agent chooses the action with the highest $Q_e(  \mathcal{S}_{t}|\mathcal{A};\bm{\theta})$ value with probability $\epsilon$, and randomly chooses an action with probability $1-\epsilon$. At the start of the training when $\epsilon$ is high, the agent explores the action space via randomly choosing the action. As $\epsilon$ decreases, the agent begins to exploit the accumulated knowledge via choosing the action with the highest $Q_e(  \mathcal{S}_{t}|\mathcal{A};\bm{\theta})$ value. The parameters of the proposed algorithm are summarized in Table~\ref{table1}.

\begin{table}[]
\centering
\caption{Algorithm hyper-parameters}
\label{table1}
\begin{tabular}{||c c||} 
\hline
Parameter                  & Value                      \\ \hline
No. of hidden layers	       & 3         \\ \hline
Discount factor $\gamma$     &0.05            \\ \hline
Learning rate $\alpha$       & 0.001                       \\ \hline
$\epsilon$                   & 1 \text{to} 0.01                          \\ \hline
\end{tabular}
\end{table}

\subsection{Performance Evaluation}

 \subsubsection{Synthetic Data}



\begin{figure}
    \centering
    \includegraphics[width=5.7in,height=9cm]{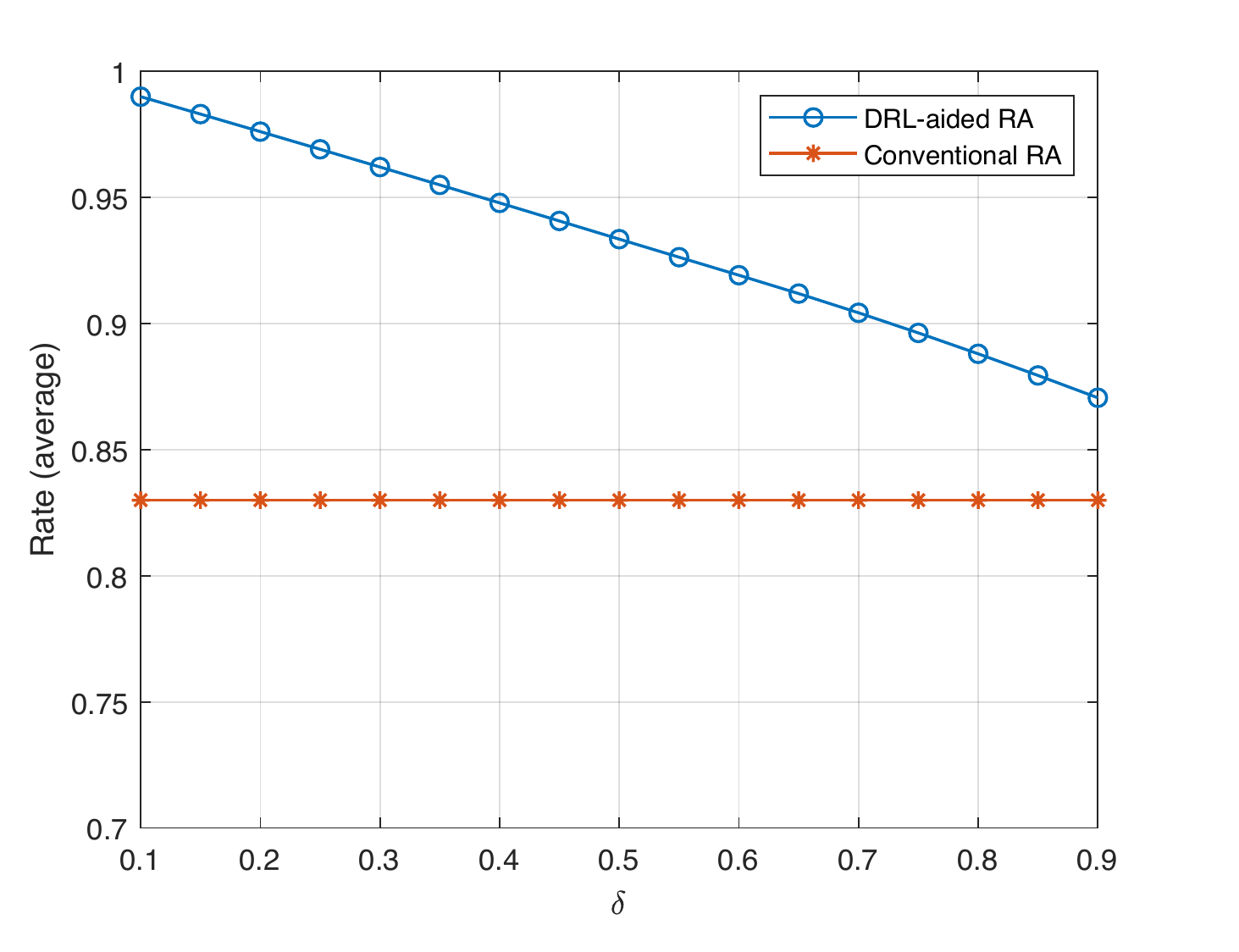}
    \caption{Average packet rate as a function of $\delta$ for synthetic activity patterns.}
    \label{fig:ac_delta}
\end{figure}

In Fig.~\ref{fig:ac_delta}, we present the average packet rate achieved with the proposed DRL-aided RA scheme on the synthetic activity pattern generated by (\ref{syn_act}) and compare it with the packet rate achieved with the conventional RA scheme for different values of $\delta$. In this example, the number of nodes in the cell is set to $K = 20$ and the number of available RBs at the AP is set to $N = 10$. The number of RBs allocated by the AP via the DRL agent decreases from $N_1 = 7$ for $\delta = 0.1$, to $N_1 = 1$ for $\delta = 0.9$. To determine $N_1$, we only need to know the probability that a node is correctly classified as active and the probability that a node is misclassified as inactive (see \eqref{prob_RA}-\eqref{m2}), which are obtained from the data samples used for training. In particular, we use these samples to calculate the rate using \eqref{prob_RA} for all values of $N_1$, and we chose the value which results with the highest rate. In the case of transfer learning, we use only the samples from the real-world activity pattern (not the data used for pre-training). The number of available orthonormal sequences is set to $M = 54$. As Fig.~\ref{fig:ac_delta} illustrates, the average packet rate of the conventional RA scheme is not sensitive to $\delta$. On the other hand, the average packet rate of the proposed DRL-aided RA scheme is a decreasing function of $\delta$. This is due to the amount of randomness in the activity patterns as $\delta$ increases. In particular, when $\delta$ is low the activity pattern is almost periodic so the DRL agent is able to learn it and then correctly allocate the available RBs, thereby reducing the need for the nodes to attempt RA. Conversely, when $\delta$ is high the activity pattern is highly random and the agent is not able to learn it completly, so the nodes attempt RA to gain RBs. This example illustrates that the average packet rate of the proposed DRL-aided RA scheme is lower bounded by the average packet rate of the RA scheme. Thereby, the worst possible performance of the proposed DRL-aided RA scheme, obtained for $\delta = 1$, is identical to the performance of the RA scheme.

\begin{figure}[tbp]
\centering
\includegraphics[width=5.7in,height=9cm]{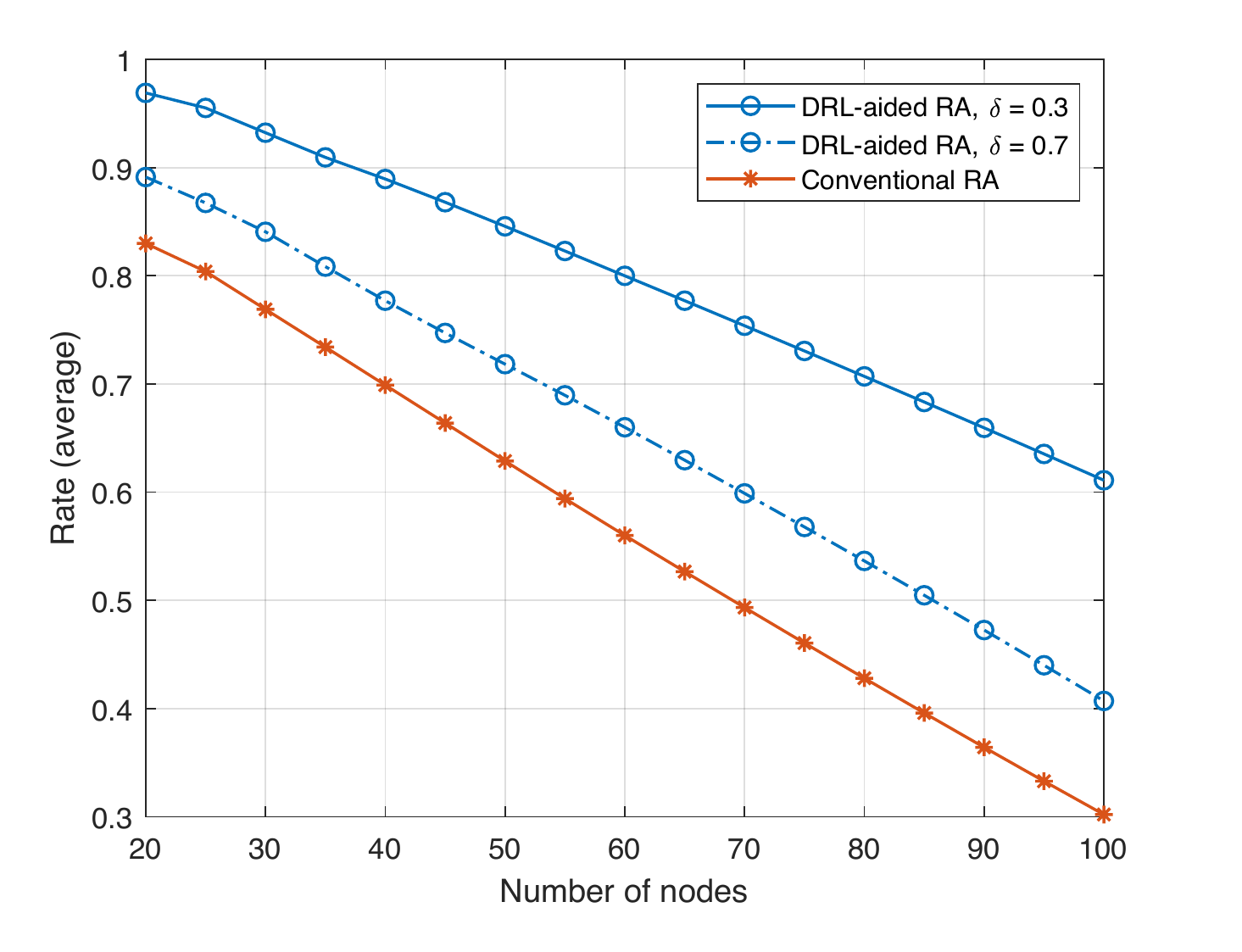}
\caption{Average packet rate as a function of the number of nodes for synthetic activity patterns.}
\label{num3_a}
\end{figure}

In Fig.~\ref{num3_a}, we illustrate average packet rate achieved with the proposed DRL-aided RA scheme and compare it with the average packet rate of the conventional RA scheme as a function of the number of nodes in the cell $K$, for two different values of $\delta$. The number of available RBs at the AP is set to $N = 10$. The number of RBs allocated by the AP via the DRL agent is set to $N_1 = 5$ and $N_1 = 2$ for $\delta = 0.3$ and $\delta = 0.7$, respectivley. The number of available orthonormal sequences is set to $M = 54$. As it can be seen from Fig.~\ref{num3_a}, the packet rate of the proposed DRL-aided RA scheme is significantly higher than the rate of the RA scheme. For example, the proposed scheme can achieve a packet rate of $0.6$ when $K = 100$ nodes, whilst the conventional RA scheme can achieve the same packet rate with $K = 50$ nodes when $\delta = 0.3$. Similarly, when the activity patterns are more random i.e., when $\delta = 0.7$, the proposed scheme can achieve a packet rate of $0.6$ for $K = 70$ nodes, whilst the conventional RA scheme achieves the same rate with $K = 50$ nodes.

\subsubsection{Real-World Activity Pattern}

\begin{figure}[tbp]
\centering
\includegraphics[width=5.5in,height=9.5cm]{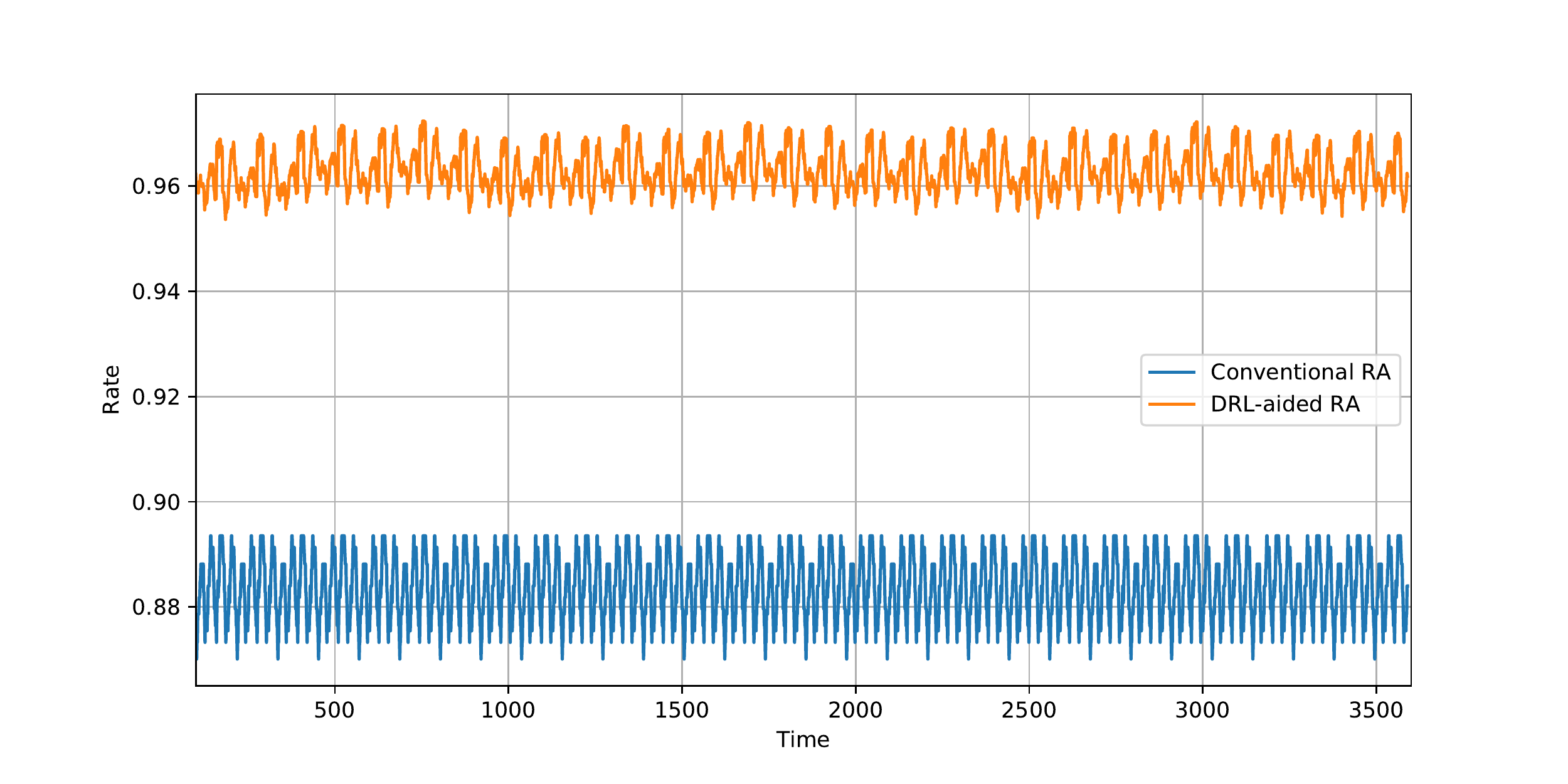}
\caption{Instantaneous packet rate in each time slot during one hour for real-world activity patterns.}
\label{num3_b}
\end{figure}

In Fig.~\ref{num3_b}, we illustrate the instantaneous packet rate  in each time slot during a period of 1 hour (3600 s) for the real-world activity pattern. In total our IoT cell is comprised of $K=222$ nodes, which report $35448$ data arrivals during one hour. The number of available RBs at the AP is set to $N = 20$. The number of RBs allocated by the AP via the DRL agent is set to $N_1 = 16$. The number of available orthonormal sequences is set to $M = 54$. Since the minimum duration between two data arrivals in these data sets is $1$s, we assume that the duration of a time slot is $t=1$s. As it can be seen from Fig.~\ref{num3_b}, the proposed DRL-aided RA scheme achieves a packet rate that is significantly higher than  the conventional RA scheme in each time slot. This is a consequence of the fact that the agent is able to extract the determinism in the activity pattern, which exists in practice, and correctly predict some of the active nodes. As a result, the number of nodes that attempt the RA procedure is much lower compared to the conventional RA scheme. In the spirit of reproducible science, the codes used for generating this figure are made available on \cite{ikoloska}.

\begin{figure}[tbp]
\centering
\includegraphics[width=6.1in,height=9cm]{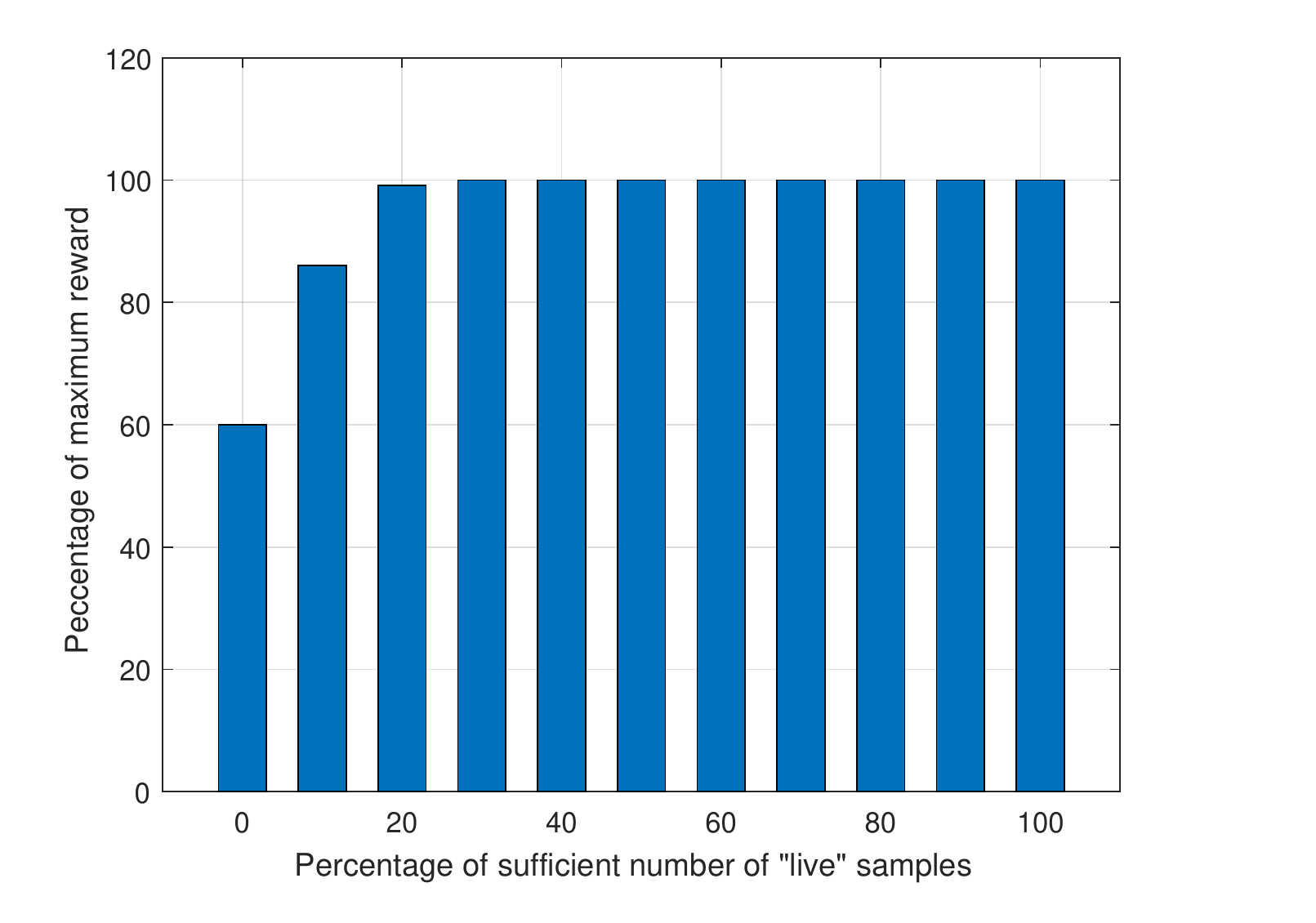}
\caption{Comparison between transfer learning and data driven learning. The percentage of artificial samples can be found as $100-X$, where $X$ denotes the percentage of sufficient number of live samples.}
\label{lTl1}
\end{figure}

To illustrate the benefits of transfer learning we present Fig.~\ref{lTl1}, where the percentage of maximum possible reward is illustrated as a function of the percentage of sufficient live samples. The sufficient number of live samples is defined as the number of live samples needed for the DRL agent to obtain the maximum possible reward, and thereby achieve the maximum possible inference accuracy. 
The maximum reward is defined as the reward obtained by using $100\%$ of live data samples. Fig.~\ref{lTl1} shows that the maximum reward can be obtained by using $20\%$ of live data samples and $80\%$ of artificial samples. Note that, using an insufficient number of live data samples, and no artificial samples, leads to a reward that is significantly lower than the maximum possible reward, as the agent does not have enough data for the training process. In addition, the obtained reward is significantly lower if only artificial samples, without any live samples, are used which is a consequence of the mismatch between the artificial model and the activity in the IoT cell. Thereby, optimal performance can be achieved by using $20\%$ of live samples and $80\%$ of artificial samples from the theoretical model. This in turn significantly decreases the time required for the agent to be trained, i.e., by up to $80 \%$ in our case.


\section{Conclusion}
In this paper, we proposed a DRL-aided RA scheme for a network comprised of $K$ IoT nodes and an AP. In particular, an intelligent DRL agent placed at the AP learns to predict the  activity of the IoT nodes in each time slot and grants  time-frequency resource blocks to the IoT  nodes predicted as active. The standard RA scheme is used as a back-up access mechanism for potentially misclassified, unseen or new nodes in the cell. In addition, we leverage expert knowledge in order to ensure faster training of the DRL agent. By using publicly available data sets, we show signifficant improvements in terms of rate, when the proposed DRL-aided RA scheme is implemented, compared to the conventional RA scheme.


\bibliographystyle{IEEEtran}
\bibliography{litdab}
\end{document}